\documentclass[preprint,3p]{elsarticle}
\usepackage{cite}
\usepackage{caption}
\usepackage[utf8]{inputenc}
\usepackage{hyperref}
\usepackage[T1]{fontenc}
\usepackage{array}
\usepackage{mathtools,amsmath,amssymb}
\usepackage{algorithm,algorithmic}
\usepackage{bm}
\usepackage{graphicx}
\usepackage{epstopdf}
\usepackage{subfigure}
\usepackage{multirow}
\usepackage{array,booktabs}
\usepackage{tabularx,framed}
\usepackage{tikz}
\usetikzlibrary{shapes,arrows,fit,positioning}

\biboptions{comma,square,sort&compress}
\usepackage{lineno,hyperref}
\modulolinenumbers[5]

\journal{Biomedical Signal Processing and Control}

\begin{document}

\begin{frontmatter}

\title{EEG-based Texture Roughness Classification in Active Tactile Exploration with Invariant Representation Learning Networks}

\author[label1,label2]{Ozan \"{O}zdenizci\corref{cor1}\fnref{fn1}}
\author[label3]{Safaa Eldeeb\fnref{fn1}}
\author[label1]{Anda\c{c} Demir}
\author[label1]{Deniz Erdo\u{g}mu\c{s}}
\author[label3]{Murat Ak\c{c}akaya}

\address[label1]{Department of Electrical and Computer Engineering, Northeastern University, Boston, MA, USA}
\address[label2]{Institute of Theoretical Computer Science, Graz University of Technology, Graz, Austria}
\address[label3]{Department of Electrical and Computer Engineering, University of Pittsburgh, Pittsburgh, PA, USA}

\cortext[cor1]{Corresponding author: oozdenizci@ece.neu.edu}
\fntext[fn1]{Equal contribution}

\begin{abstract}During daily activities, humans use their hands to grasp surrounding objects and perceive sensory information which are also employed for perceptual and motor goals. Multiple cortical brain regions are known to be responsible for sensory recognition, perception and motor execution during sensorimotor processing. While various research studies particularly focus on the domain of human sensorimotor control, the relation and processing between motor execution and sensory processing is not yet fully understood. Main goal of our work is to discriminate textured surfaces varying in their roughness levels during active tactile exploration using simultaneously recorded electroencephalogram (EEG) data, while minimizing the variance of distinct motor exploration movement patterns. We perform an experimental study with eight healthy participants who were instructed to use the tip of their dominant hand index finger while rubbing or tapping three different textured surfaces with varying levels of roughness. We use an adversarial invariant representation learning neural network architecture that performs EEG-based classification of different textured surfaces, while simultaneously minimizing the discriminability of motor movement conditions (i.e., rub or tap). Results show that the proposed approach can discriminate between three different textured surfaces with accuracies up to 70\%, while suppressing movement related variability from learned representations.\end{abstract}


\end{frontmatter}


\section{Introduction}

Active tactile exploration refers to the process of exploring surrounding objects to retrieve sensory information \citep{lederman2009haptic}. Dynamic active movement between human skin and an object's surface may occur during this process. Active tactile exploration involves several parts of the human body which are characterized by high density sensory receptors, high sensory acuity and large sensory and motor cortical representation \citep{gibson1962observations}. The high density mechanoreceptors on the human skin contribute significantly to the discrimination between different objects varying in pattern, texture and shape \citep{borich2015understanding,abraira2013sensory}. Mechanoreceptors in the epidermis and dermis layers of the human fingertips include distinct small, large, slow and rapidly adapting receptive fields which enable high spatial resolution \citep{borich2015understanding,abraira2013sensory}. Active tactile object exploration generates simultaneous cutaneous and proprioceptive feedbacks, whose combination is referred as haptic feedback. Proprioceptive signals originate from joint, muscle and skin mechanoreceptors, and are related to joint movement and position \citep{gibson1962observations}.

The primary somatosensory cortex (S1) is known to be the main brain region responsible for processing sensory information related to active exploration of objects \citep{borich2015understanding}. This region is also known to have a direct association with the primary motor cortex, which is also involved in processing of active exploratory motor movements. During active tactile exploration, S1 cortex demonstrates neural activity that reflects the activation of several peripheral mechanoreceptors, as well as joint movements \citep{chapman1994active}. Moreover, movements associated with tactile exploration modulates the transmission of the tactile input to S1 cortex, adding further complications to the processing of somatic sensory signals \citep{chapman1994active}. It has also been shown that a vast majority of neural activity in S1 cortex significantly contributes to the processing of tactile input during active tactile exploration compared to passive exploration \citep{blatow2007fmri,singh2014brain}. Accordingly, active tactile exploration provides substantial information about the surface properties and demonstrate better discrimination thresholds compared to passive exploration \citep{lederman2009haptic,gibson1962observations,hollins2000evidence}.

Sensorimotor processing is generally visualized as a sequence of function units that are sequentially activated that initiates from sensory input and proceeds until motor execution \citep{shadlen2001neural,sugrue2005choosing,gold2007neural}. Several other research studies on human brain function suggest that motor and sensory processing are coupled, where motor actions and perception are in affect with each other \citep{engel2013s,konig2013predictions,gallese2005brain}. A study by Melink et. al. \citep{melnik2017eeg}, investigated these two views of sensorimotor processing (i.e., the classical view and the alternative coupling perspective). Their electroencephalography (EEG) based exploratory study supports the alternative sensorimotor processing hypothesis that sensory perception and motor responses are coupled.

Research studies investigating EEG-based cortical activity in response to active tactile exploration are generally explored within passive touch experimental designs \citep{genna2016long,moungou2016eeg,genna2018bilateral}. To the best of our knowledge, there exists only a few studies which explores the relation between recorded EEG and the roughness of the explored surface in active touch \citep{moungou2016novel,eldeeb2019eeg}. Yet, EEG-based classification of textured surfaces that vary in their roughness levels, while minimizing the potential influence of motor movement type during active tactile exploration remains to be explored.

In this study, we aim to classify different textured surfaces during active tactile exploration that vary in the roughness level on a single trial basis using simultaneously recorded EEG data. Our experimental study design includes two motor movement conditions, rubbing or tapping the textured surface, and three different levels of surface roughness (i.e., smooth, medium rough and rough). We propose an invariant representation learning neural network that allows classification of textured surface roughness levels, while minimizing the discriminability of the motor movement type in a systematic way using an adversarial training approach \citep{ozdenizci2020learning}. The overarching goal of this study is to use such a framework to develop a system that mimics the sensation of surfaces with varying levels of roughness during active exploration. These systems would ideally be used in a wide range of applications such as teleoperation, surgical training of physicians in virtual environments, or remote control of robotic prostheses \citep{Melchiorri2013,5152670,4479950,7161354,Pacchierotti:2015}.


\section{Experimental Study Design}

\subsection{Participants}

A total of eight right-handed healthy participants (age: 26 $\pm$ 3.4 years) participated in this study. Participants did not have any neurological or somatosensory deficits or physical limitations. Written informed consents (University of Pittsburgh IRB No. PRO17040151) were obtained from all participants.

\subsection{Experimental Setup}

\begin{figure}[t!]
    \centering
    \subfigure[]{\includegraphics[height=5.5cm]{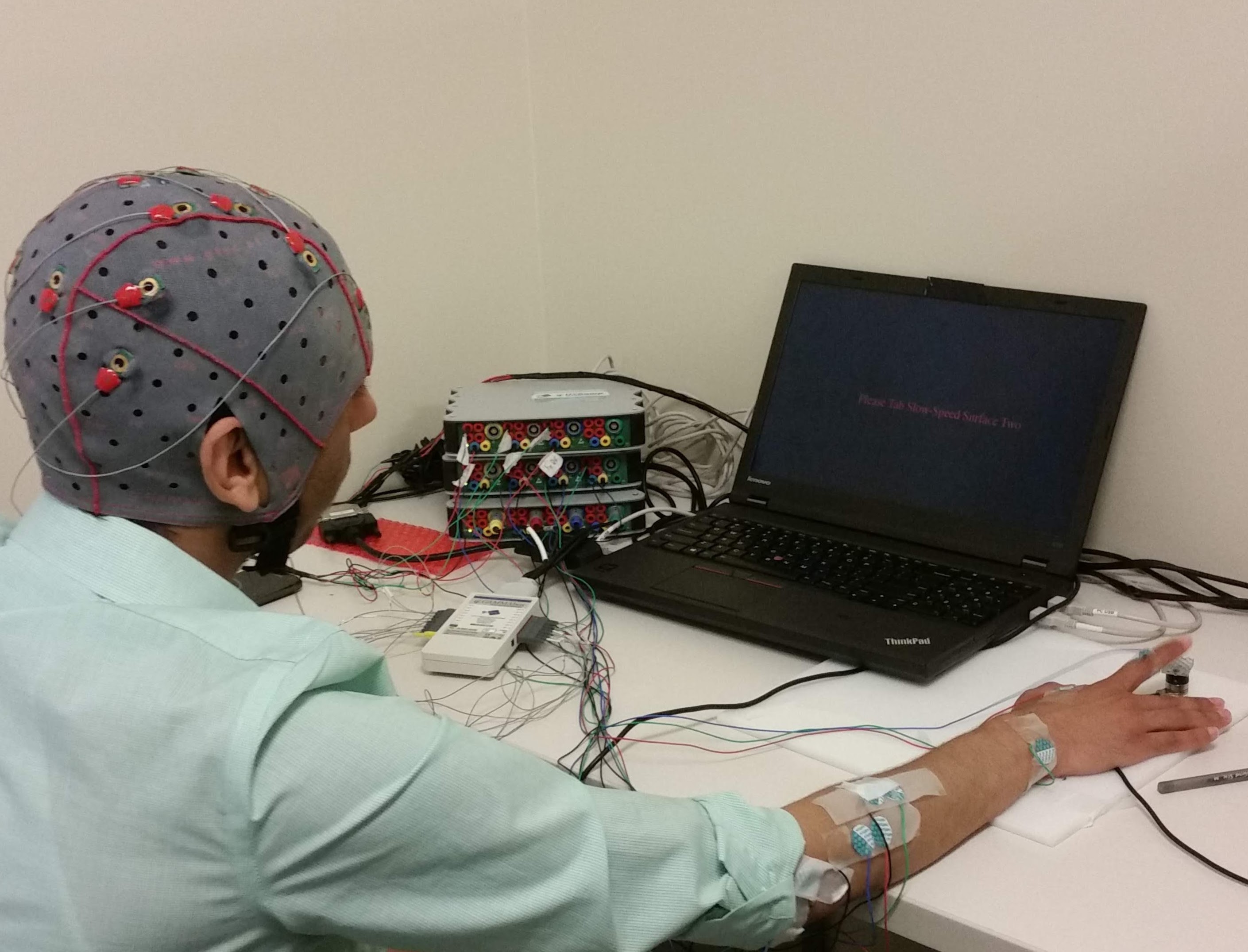}\label{fig:exp1}}\hspace{1cm}
    \subfigure[]{\includegraphics[height=5.5cm]{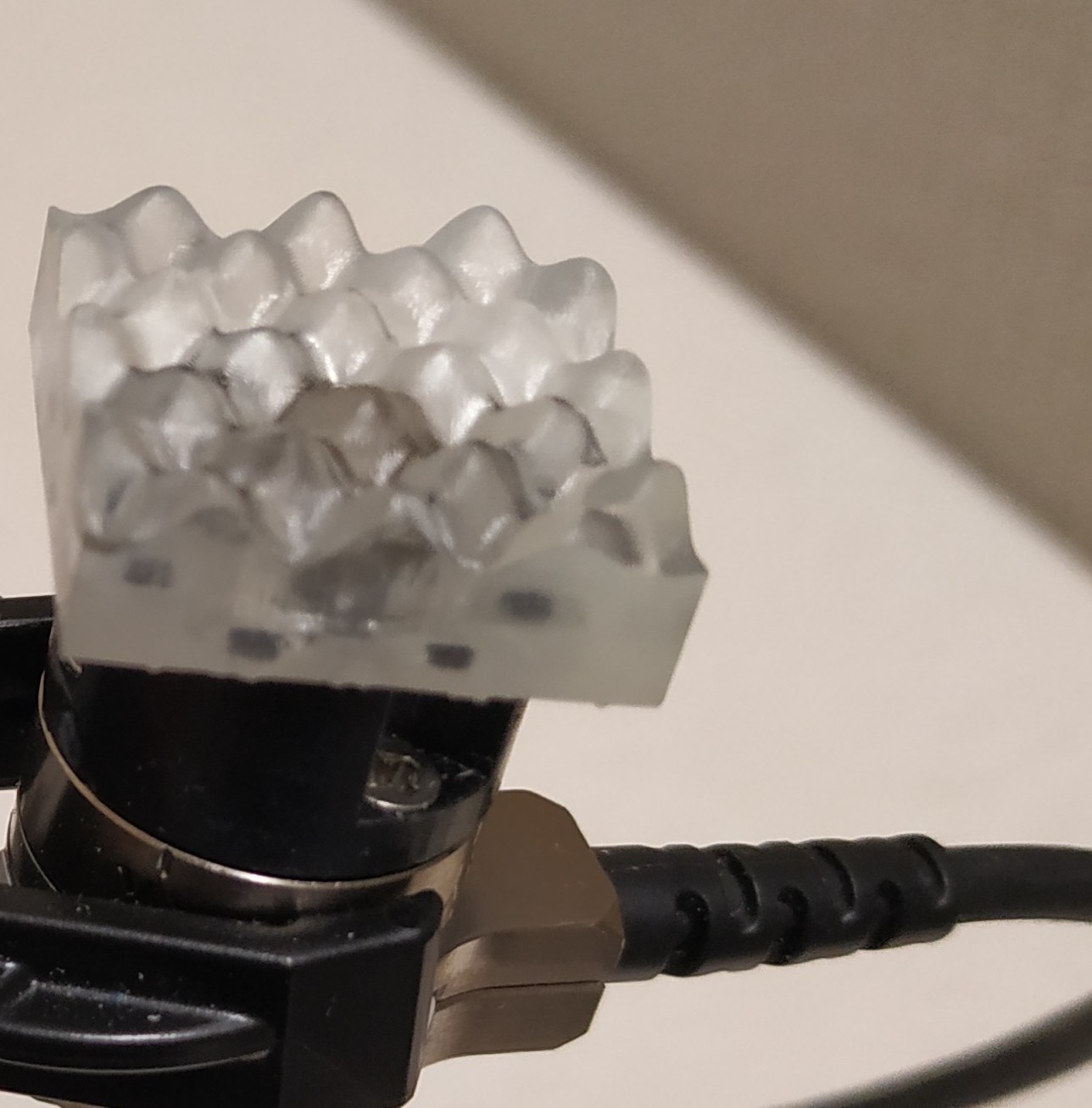}\label{fig:exp2}}
    \caption{(a) Illustration of the experimental setup during a trial where the participant is instructed to tap on the texture mounted on the force transducer. (b) Printed medium rough 3D texture mounted on the force transducer.}
    \label{fig:ExpSetup}
\end{figure}

During the experiments the participants sat in front of a computer screen and were instructed to use the tip of their dominant hand index finger while rubbing or tapping three different textured surfaces with varying levels of roughness. An illustration of the experimental setup is provided in Figure~\ref{fig:exp1} (cf.~\citep{eldeeb2019eeg} for further information regarding the experimental setup design). The computer screen provided visual cues to the participants via a graphical user interface we developed using the Psychtoolbox software \citep{brainard1997psychophysics}. These cues marked the beginning of each condition. The experiment consisted of a total of six conditions, two movement type (i.e., rub and tap); and three textures surfaces (i.e., flat smooth, medium rough and rough). For each condition we instructed each participant to rub or tap the selected surface. Each complete movement across the textured surface was considered a single complete trial. 

Three synthesized textured surfaces with varying levels of roughness (i.e., smooth flat, medium rough, and rough) were generated using MATLAB (MathWorks, USA) and fabricated with Stereolithography (Viper SLA System, 3Dsystems, USA). The smooth flat surface is an even regular surface without any form of roughness. The roughness of the medium rough and rough surfaces represented by the power spectral density was controlled by the following expression: 
$$
\phi(|k|) = \begin{cases}
    C, & \text{if $k_l<= |k| <= k_r$}.\\
     C (\frac{|k|}{k_r})^{-2(1+H)}, & \text{if $k_r <= |k| <= k_s$}.\\
    0, & \text{otherwise}.
  \end{cases} \eqno{(1)}
$$
where $C$ is the roughness amplitude,  $k_l$, $k_r$, $k_s$ are the lower roll-off and upper cutoff wave numbers and $H$ is the Hurst roughness exponent. The values for each surface in this study are chosen as follows: the medium rough surface (H = 0.5, C = $10*10^{10}$, $k_l$ = $k_r$ = 16, $k_s$ = 64) and rough surface (H = 0.5, C = $10*10^{10}$, $k_l$ = $k_r$ = 32, $k_s$ = 256). For the third texture, simply a smooth flat surface was used. Each synthetic textured surface was mounted on a force transducer (as illustrated in Figure~\ref{fig:exp2} for the medium rough texture), and adjusted on a table in reach of the participants.

\subsection{EEG Data Acquisition}

Throughout the experiments, EEG data were collected from 14 EEG channels placed over the frontal and somatosensory cortices. These channels were placed according to the international 10-20 system \citep{klem1999ten} as channel locations: F3, F4, FC3, FC4, C1, C3, C5, CZ, C2, C4, C6, CP1, CPZ and CP2. We used the left mastoid as a reference and FPz as the ground electrode. Force data generated from touching the textured surfaces were also recorded using a force and torque transducer (NANO17 F/T transducer, ATI Industrial Automation, USA). Two g.USBamp amplifiers (g.tec medical engineering GmbH, Graz, Austria) were used to record and synchronize both EEG and force data. The first amplifier, which is used for EEG data acquisition, has a sampling frequency of 1200 Hz and uses a 4th order notch filter with cutoff frequencies 58 and 62 Hz, and an 8th order bandpass filter with cutoff frequencies of 2 and 62 Hz. The second amplifier was connected to the force transducer and used to synchronize the collected force data and digitize it with sampling rate of 1200 Hz. The force data was used to mark the beginning of each trial for the continuously recorded EEG data. We downsampled EEG data to 300 Hz sampling rate offline, and segmented into trials with duration of one third of a second following the touch (i.e., $\sim$330 ms).

\section{Methods}

Deep neural networks have been widely explored as generic feature extractors for EEG in various classification tasks \citep{Bashivan:2016,schirrmeister2017deep,lawhern2018eegnet,fahimi2019inter,Craik:2019}. We recently explored such discriminative models in the context of deep invariant representation learning \citep{xie2017controllable,lample2017fader,moyer2018invariant}, to illustrate potential capabilities of EEG-based neural network models to censor specific nuisance information inherent within the learned feature extractors \citep{ozdenizci2019ner,ozdenizci2019adversarial,ozdenizci2020learning}. Our proposed neural network aims to learn texture roughness discriminative features while minimizing the influence of movement conditions (rub or tap) during active tactile exploration. We realize this using an adversarial training approach which infers a nuisance (i.e., motor movement activity) invariant latent space within a discriminative setting, hence suppressing motor movement related cortical activity from representations that are capable of classifying texture roughness levels.

\subsection{Invariant Representation Learning Network}

\begin{figure*}[t!]
    \centering
    \includegraphics[width=0.94\textwidth]{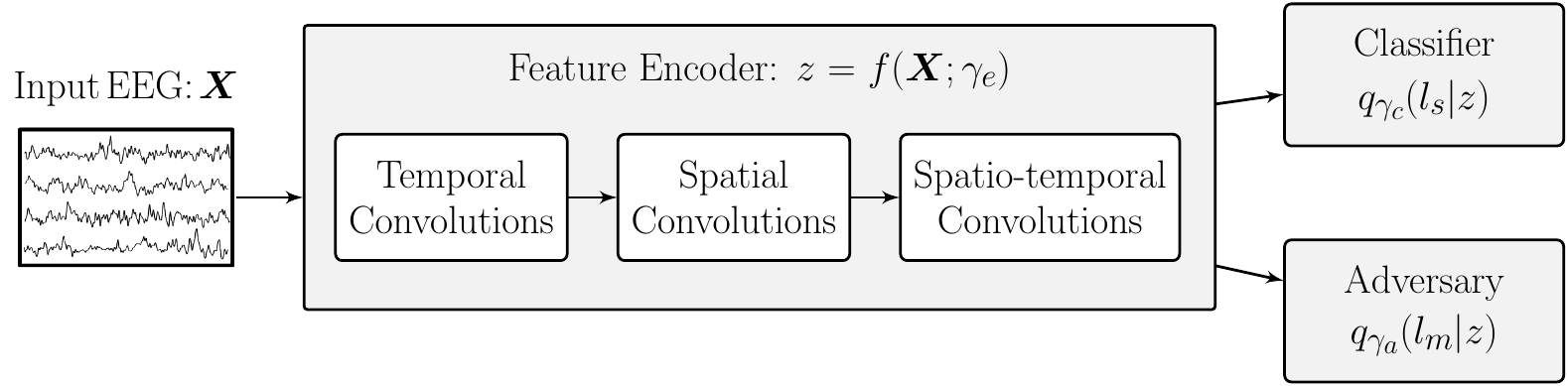}
    \caption{Overall adversarial invariant representation learning network flow, illustrating the feature encoder, classifier and adversary networks. Network layer specifications are further provided in Table~\ref{tab:network}.}
    \label{fig:architecture}
\end{figure*}

We consider a subject-specific EEG data set $\{(\mathbf{X}_{i}, {l_{s}}_{i}, {l_{m}}_{i})\}_{i=1}^{n}$ with $n$ number of trials, where $\mathbf{X}_{i} \in \mathbb{R}^{CxT}$ denotes the multichannel EEG data collected from $C$ channels and $T$ number of discretized time samples, ${l_{s}}_{i}$ denotes the label for the roughness of a texture surface, and ${l_{m}}_{i}$ denotes the binary nuisance variable (i.e., rub or tap motor movement). Since no ordinal relationship exists in the categorical variables ${l_{s}}$ and ${l_{m}}$, we represent them as one-hot encoded label vectors. Nuisance variations denoted by $l_{m}$ are assumed to be involved in the data generation process, and therefore we employ an adversarial training approach that can discriminate $l_{s}$ invariant of $l_{m}$. We premise the underlying data generation process on our assumption that ${l_{s}}$ and ${l_{m}}$ are disentangled such that $l_{s} \sim p(l_{s})$ and $l_{m} \sim p(l_{m})$ with $\mathbf{X} \sim p(\mathbf{X}|l_{s}, l_{m})$.

The network flow is outlined in Figure \ref{fig:architecture}. A deterministic feature \textit{encoder network} with parameters $\gamma_{e}$ projects the time-series EEG data onto a lower dimensional latent feature space, $z=f(\mathbf{X};\gamma_{e})$. This latent vector is passed to a \textit{classifier network} with parameters $\gamma_{c}$, and an \textit{adversary network} with parameters $\gamma_{a}$ separately. The goal of classifier network is to infer surface texture label $l_{s}$, whereas that of adversary is to infer the nuisance variations $l_{m}$ evoked by motor activity. In other words, the classifier network is trained to maximize the likelihood $q_{\gamma_{c}}(l_{s}|z)$, and the adversary network maximizes its likelihood $q_{\gamma_{a}}(l_{m}|z)$. However, the joint feature encoder network parameters are trained using a loss function that combines these two objectives within a conflicting relationship, by maximization of the likelihood $q_{\gamma_{c}}(l_{s}|z)$ and minimization of the likelihood $q_{\gamma_{a}}(l_{m}|z)$. This adversarial training behavior of the adversary network enables learning of $l_{m}$-invariant feature representations. Overall, the joint feature encoder aims to learn $l_{s}$-discriminative features, while the antagonistic adversarial loss enables concealing of $l_{m}$-relevant information from the feature encoder. This corresponds to the following min-max objective for adversarial model training:
\begin{equation}
    \begin{split}
        \min_{\gamma_{e}, \gamma_{c}} \max_{\gamma_{a}}
        \mathbb{E}[-\log {q_{\gamma_{c}}}(l_{s}|f(\mathbf{X};\gamma_{e})) + \lambda\log {q_{\gamma_{a}}}(l_{m}|f(\mathbf{X};\gamma_{e}))],
    \end{split}
    \label{eqn:objective}
\end{equation}
to obtain the optimal network parameters $\hat{\gamma}_{e}, \hat{\gamma}_{c}, \hat{\gamma}_{a}$, where $\lambda>0$ is the adversarial regularization weight for nuisance-invariant decoding. $\lambda=0$ would correspond to a traditional neural network training procedure without any manipulation. In case $\lambda<0$, the encoder would be trained to learn particularly motor movement dependent features for texture roughness classification, which is not the intention of this study. Parameter updates are performed alternatingly. At each iteration, firstly the loss of the adversary is back-propagated to update the parameters of the adversary network $\gamma_{a}$ towards maximizing its log-likelihood (i.e., max objective). Afterwards, the overall loss is back-propagated to update the encoder and classifier parameters $\gamma_{e}$ and $\gamma_{c}$ (i.e., min objective). Algorithm~\ref{alg:training} outlines the stochastic model training process \citep{ozdenizci2020learning}. Implementations for the invariant EEG representation learning networks are publicly available at: https://github.com/oozdenizci/AdversarialEEGDecoding.

\begin{algorithm}[t!]
\caption{Adversarial invariant representation learning network training \citep{ozdenizci2020learning}}
\label{alg:training}
\begin{algorithmic}[1]
\renewcommand{\algorithmicrequire}{\textbf{Input:}}
\renewcommand{\algorithmicensure}{\textbf{Output:}}
\REQUIRE $\{(\mathbf{X}_{i}, {l_{s}}_{i}, {l_{m}}_{i})\}_{i=1}^{n}$, $\lambda>0$
\ENSURE  $\hat{\gamma}_{e}, \hat{\gamma}_{c}, \hat{\gamma}_{a}$
 \STATE Randomly select initial parameters: $\gamma_e,\gamma_c,\gamma_a$
 \FOR {$\# epochs$}
 \FOR {$\# minibatches$}
 \STATE Sample a minibatch: $\{(\bm{X}_b,{l_{s}}_{b},{l_{m}}_{b})\}_{b=1}^{B}$
 \STATE Update ${\gamma}_{a}$ with stochastic gradient ascent by: $\nabla_{\gamma_{a}} \sum_{b=1}^{B} \lambda \log q_{\gamma_a}({l_{m}}_{b} \vert z_b=f(\bm{X}_{b};\gamma_{e}))$
 \STATE Update $\gamma_{e},\gamma_{c}$ with stochastic gradient descent by: $\nabla_{\gamma_{e},\gamma_{c}} \sum_{b=1}^B [-\log q_{\gamma_c}({l_{s}}_{b} \vert z_b) + \lambda \log q_{\gamma_a}({l_{m}}_{b} \vert z_b)]$
 \ENDFOR
 \ENDFOR
\end{algorithmic}
\end{algorithm}

\subsection{Implementation and Evaluation}
\label{sec:implementation}

Table~\ref{tab:network} outlines the neural network architecture specifications that we used for EEG-based texture roughness classification. We determined the feature encoder layers in accordance with the EEGNet convolutional neural network (CNN) model \citep{lawhern2018eegnet}. Network input EEG data has dimensionality of 14 channels by 100 discretized time samples (i.e., 1/3 second segments of signals with 300 Hz sampling rate). The feature encoder network sequentially performs: (1) temporal convolutions that resemble to frequency bandpass filtering operations, (2) depthwise spatial convolutions \citep{Chollet:2017} to extract bandpass specific information from distinct cortical sources over all EEG sensors, and (3) spatio-temporal separable convolutions \citep{Chollet:2017} for summarization of all extracted information. Differently from the original EEGNet model \citep{lawhern2018eegnet}, we re-selected the temporal and spatial convolution kernel lengths in consistency with our data set structure. Initial temporal convolution kernel size is chosen as 75, indicating that as low as 4 Hz of periodic signal activity will be captured from the input signals sampled at 300 Hz. Spatial convolution kernel length was determined as 14 by the number of EEG channels. All temporal convolution operations involved zero padding to keep the temporal length same. None of the convolution operations had a bias term. We used batch normalization after convolution operations \citep{Ioffe:2015}, and dropout layers \citep{Srivastava:2014} with $p=0.5$ after average pooling operations.

Representations $z=f(\mathbf{X};\gamma_{e})$ that are learned at the output of the feature encoder are provided as input to two separate single dense layer networks (i.e., classifier and adversary). Both of these decision making layers perform linear classification. Classifier performs 3-class decoding of surface texture roughness, and adversary performs binary classification of motor movement type (rub or tap).

Evaluations were performed on a within-subject cross-validation basis by learning eight user-specific invariant representation learning neural networks. For each subject, we considered equal number of trials stratified by labels across six class conditions (i.e., two movement types and three textures), with total data set sample sizes of 2190, 426, 768, 438, 570, 762, 678, 708 for subjects 1 to 8 respectively. We generated the user-specific model training, validation, and test sets with randomly selected 70\%, 10\% and 20\% portions of the trials available from each subject. The 20\% test set split is specifically generated in 5-folds and repeated 10 times. Hence, all subject-specific classification analyses are performed by 10x5-fold cross-validation. 

Networks were trained with a minibatch size of 40 training trials for at most 500 epochs. Early training stopping was performed if the classifier loss of the validation set did not reach a new lowest value for 10 consecutive epochs. We considered the model parameters which resulted in the lowest validation loss after completion of model training. Adam optimizer \citep{Kingma:2015} was used for parameter updates once per batch. The overall network consisted of a total number of 1,661 trainable parameters. We implemented the adversarial neural network training protocol using the Tensorflow deep learning library \citep{Abadi:2016} with the Keras API \citep{Chollet:2015}.

\begin{table*}[t!]
	\caption{Invariant representation learning neural network architecture specifications.}
	\label{tab:network}
    \centering
	\begin{tabular}{l l l l l l}
		\hline
		& \textbf{Network Layer} & \textbf{Filters $\times$ (Kernel Size)} & \textbf{Output Dim.} & \textbf{\# of Parameters} \\
		\hline
		\parbox[t]{.7mm}{\multirow{11}{*}{\rotatebox[origin=c]{90}{\textbf{Feature Encoder}}}} & Input EEG: $\mathbf{X}$ &  & $(1,14,100)$ \\
		& Conv2D & $8 \times (1,75)$ & $(8,14,100)$ & 600 \\
		& BatchNorm + ELU &  & $(8,14,100)$ & 16 \\
		& Depthwise Conv2D & $2 \times (14,1)$ & $(16,1,100)$ & 224 \\
		& BatchNorm + ELU &  & $(16,1,100)$ & 32 \\
		& Average Pooling & $(1,4)$ & $(16,1,25)$ &  \\
		& Dropout ($p=0.5$) & & $(16,1,25)$ & \\
		& Separable Conv2D & $16 \times (1,16)$ & $(16,1,25)$ & 512 \\
		& BatchNorm + ELU &  & $(16,1,25)$ & 32 \\
		& Average Pooling & $(1,8)$ & $(16,1,3)$ & \\
		& Dropout ($p=0.5$) + Flatten & & $(1,48)$ & \\
		\hline
		& Dense + Softmax & $(48,3)$ & $(1,3)$ & 147 \\
		& Classifier Output: ${\hat{l}_{s}}$ & & $(1,3)$ & \\
	    \hline
	    & Dense + Softmax & $(48,2)$ & $(1,2)$ & 98 \\
		& Adversary Output: ${\hat{l}_{m}}$ & & $(1,2)$ & \\
	    \hline
	\end{tabular}
\end{table*}


\section{Results}

\begin{figure*}[t!]
    \centering
	\subfigure[Subject 1]{\includegraphics[width=0.24\textwidth]{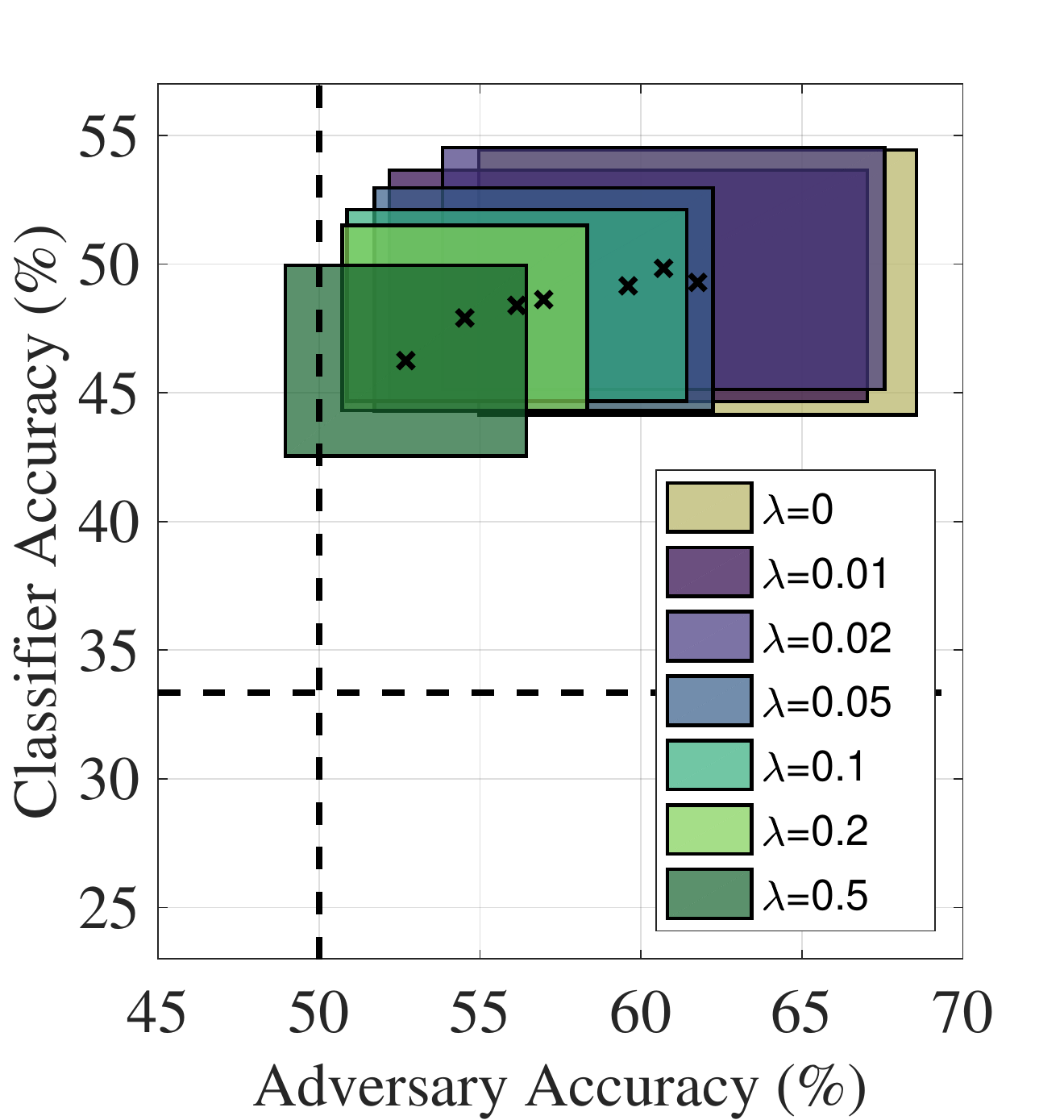}}
	\subfigure[Subject 2]{\includegraphics[width=0.24\textwidth]{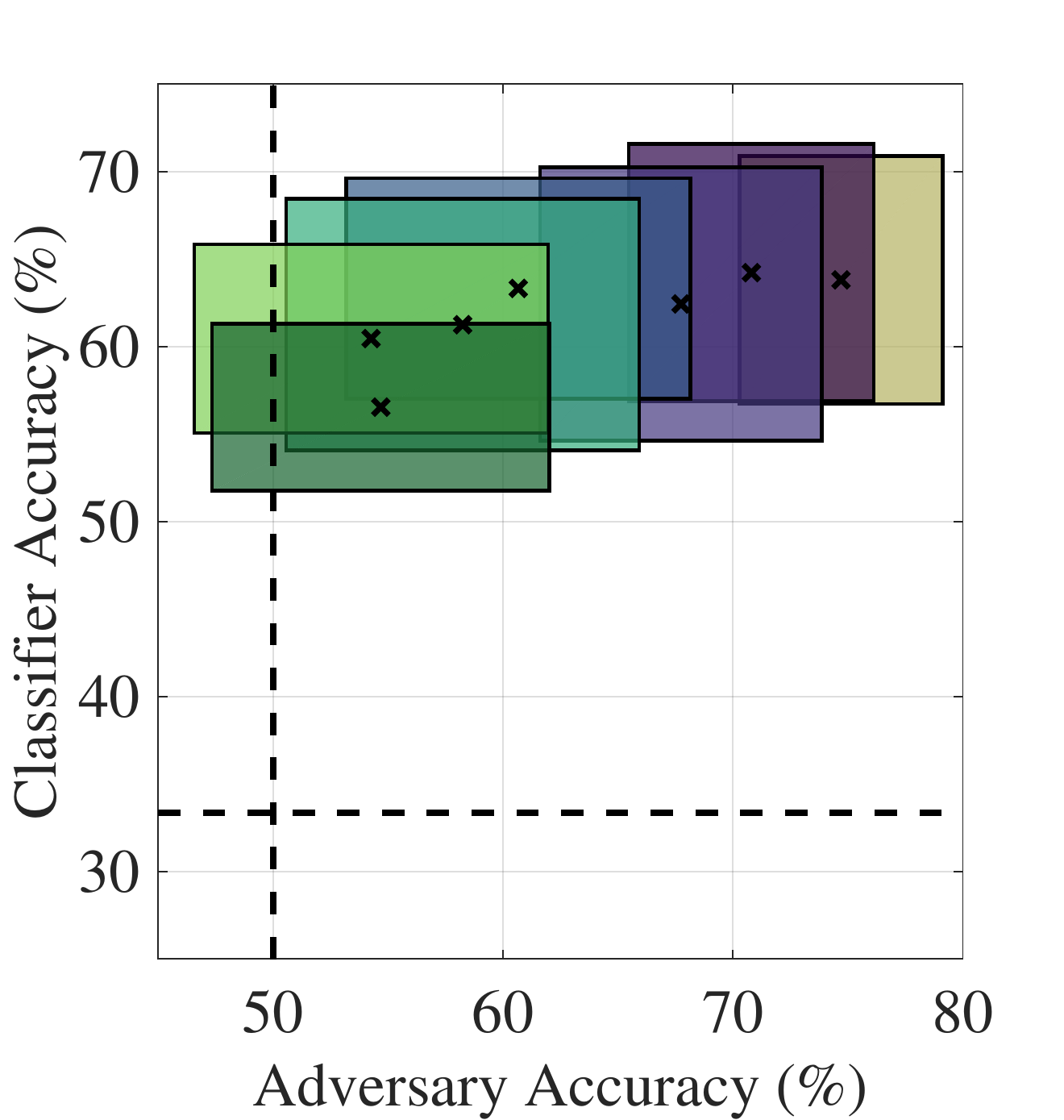}}
	\subfigure[Subject 3]{\includegraphics[width=0.24\textwidth]{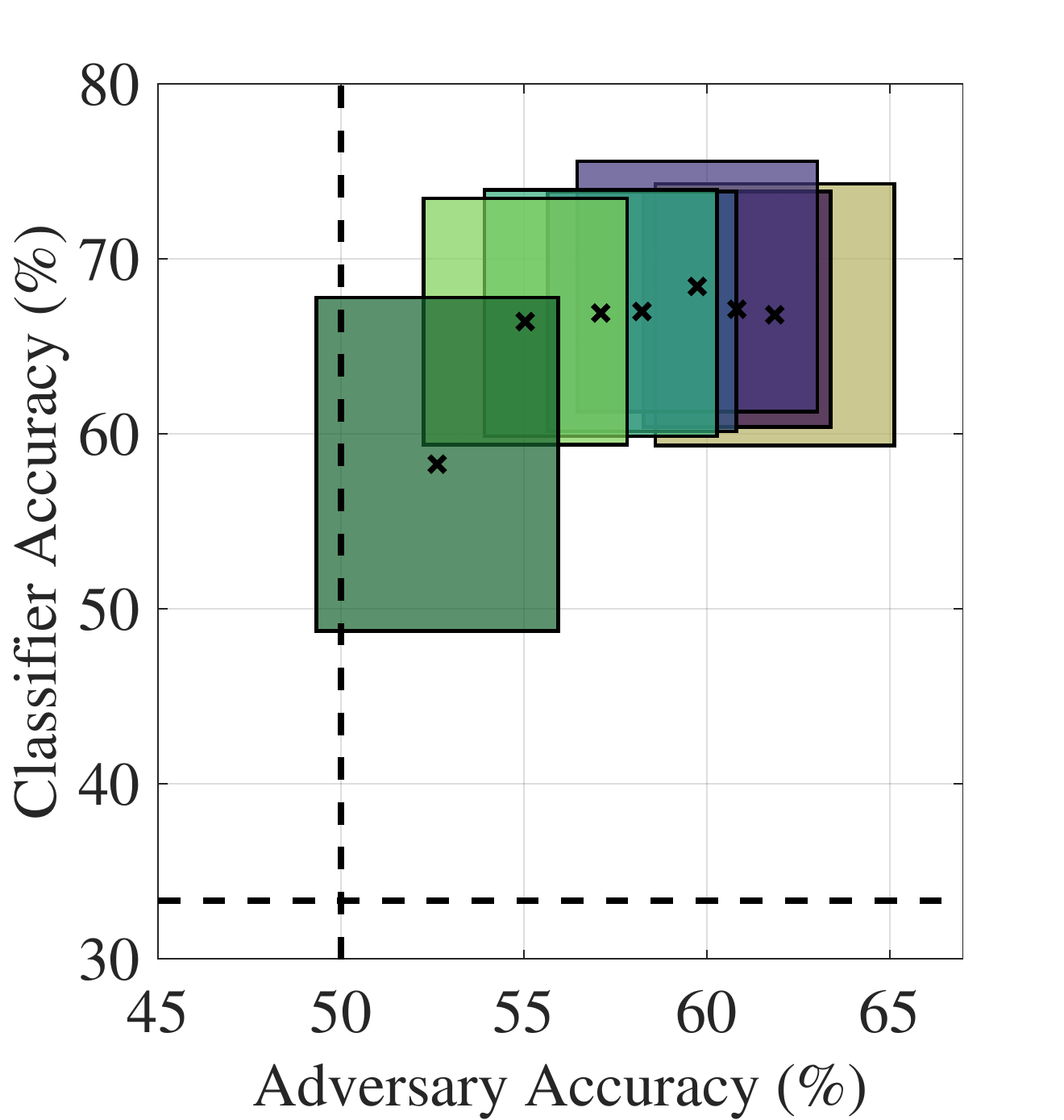}}
	\subfigure[Subject 4]{\includegraphics[width=0.24\textwidth]{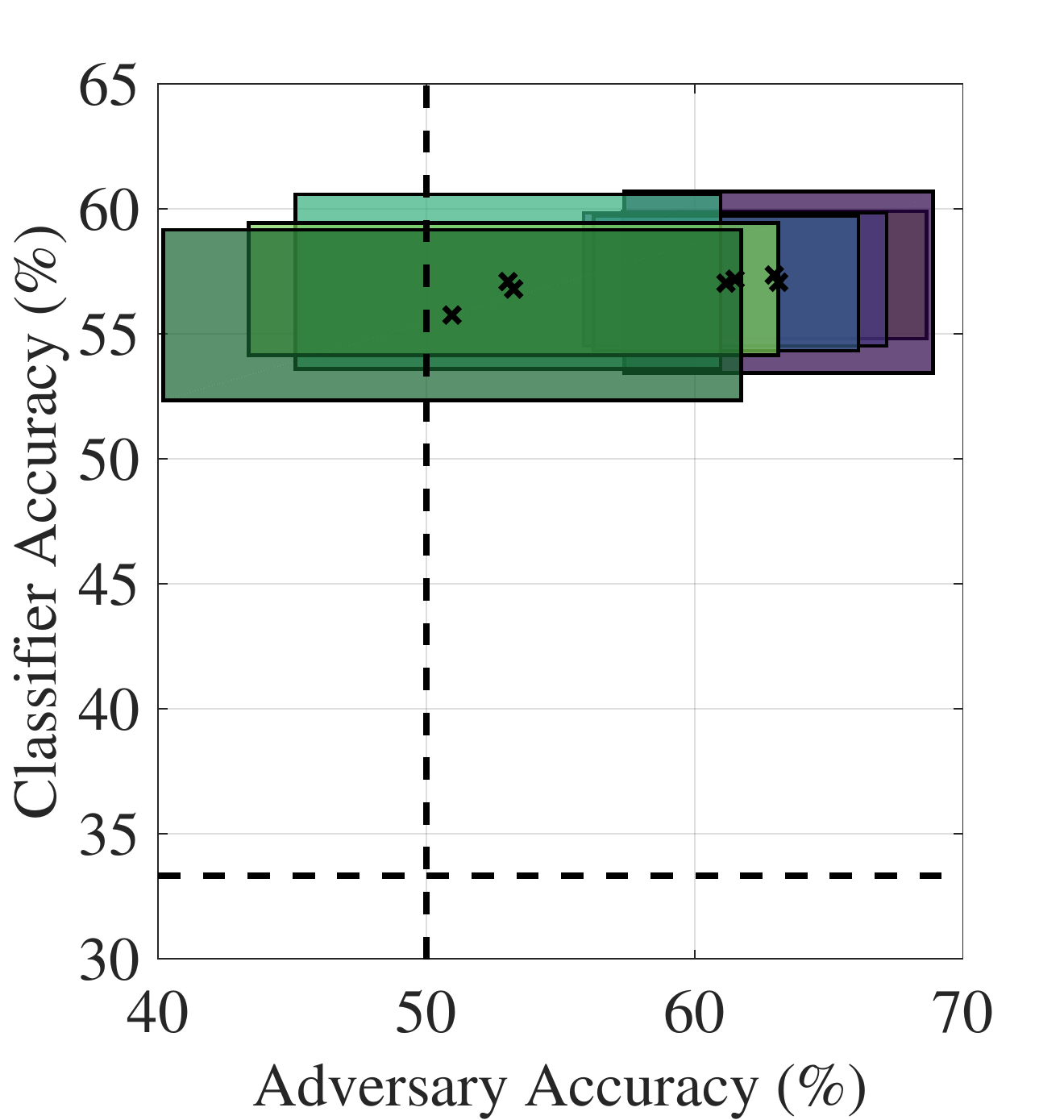}}\\
	\subfigure[Subject 5]{\includegraphics[width=0.24\textwidth]{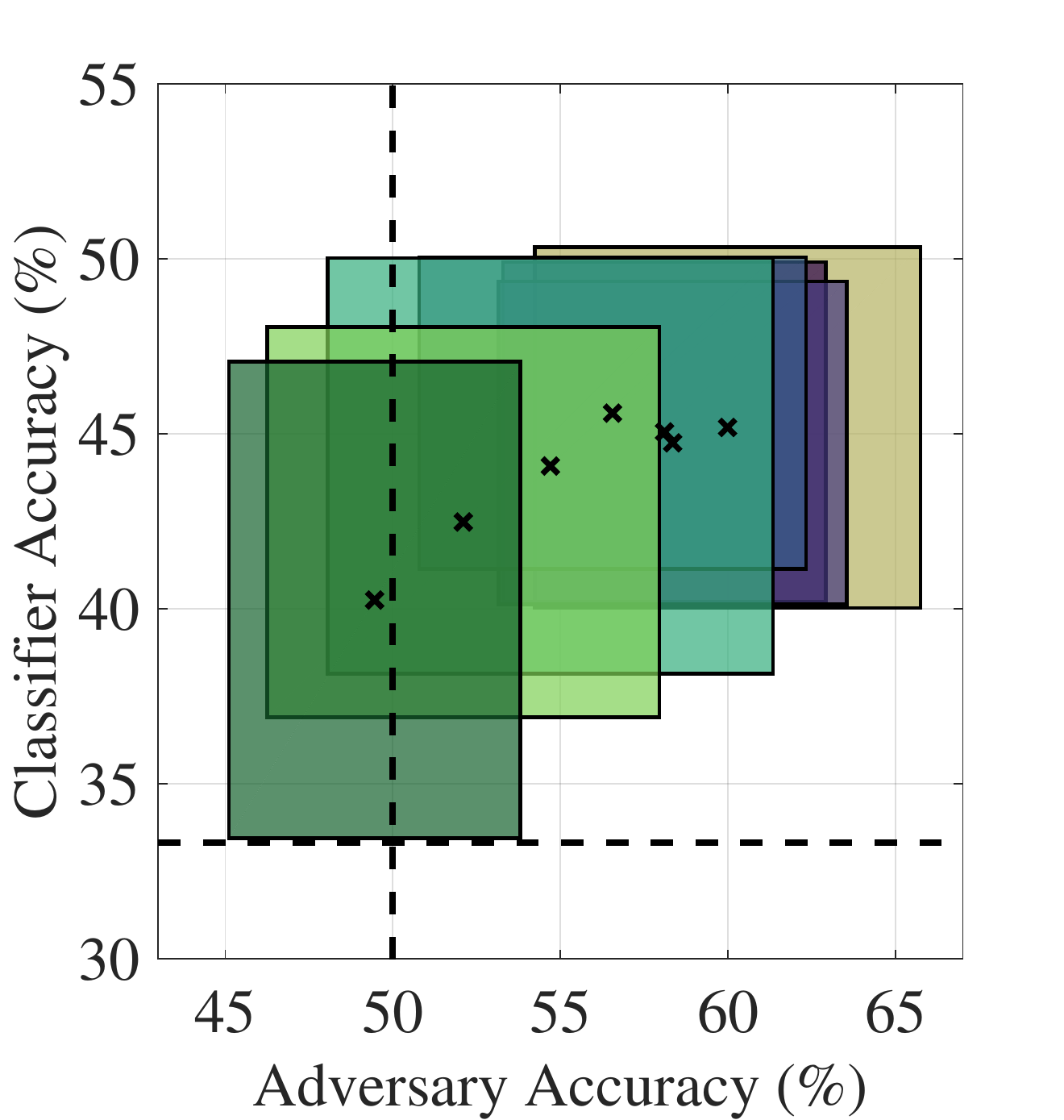}}
	\subfigure[Subject 6]{\includegraphics[width=0.24\textwidth]{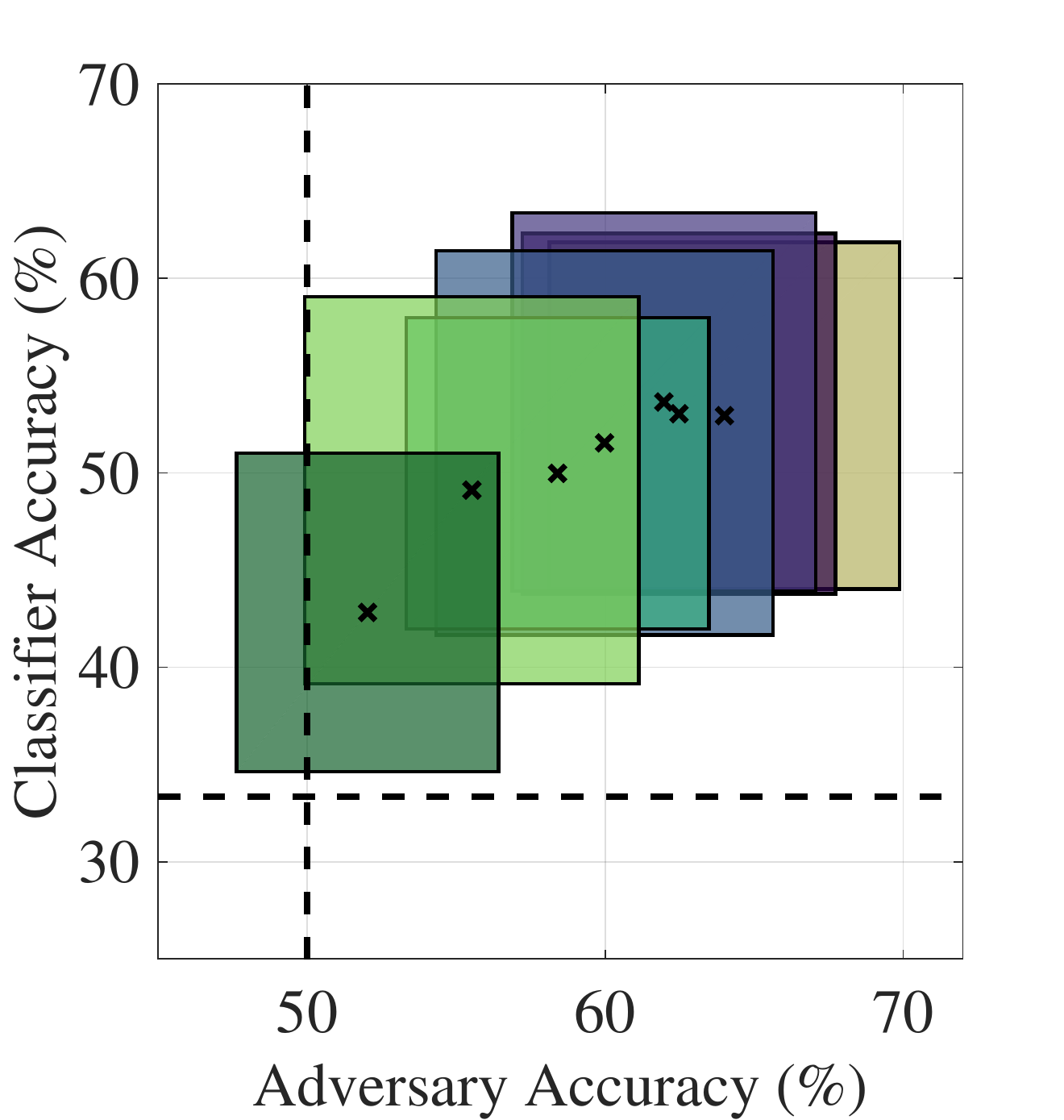}}
	\subfigure[Subject 7]{\includegraphics[width=0.24\textwidth]{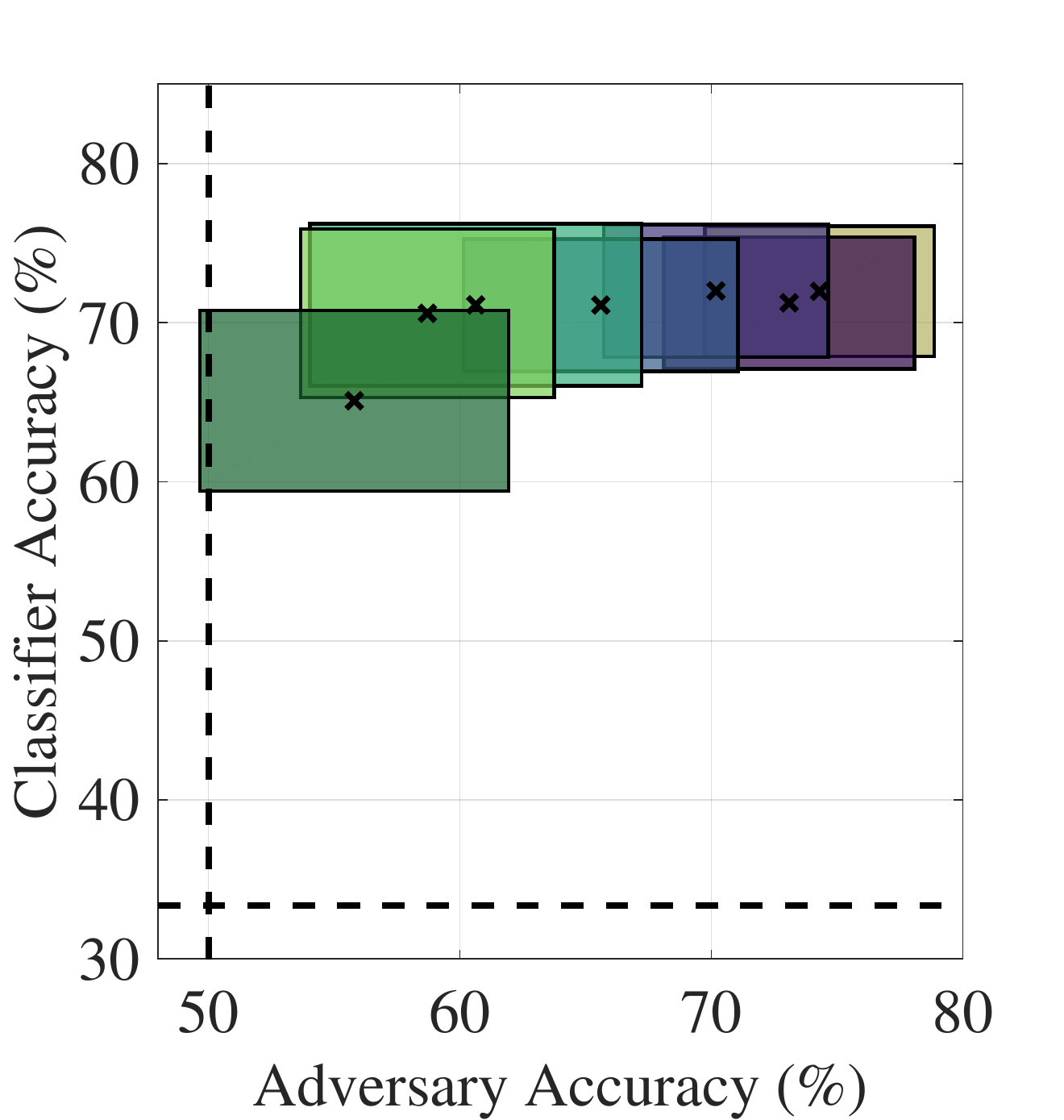}}
	\subfigure[Subject 8]{\includegraphics[width=0.24\textwidth]{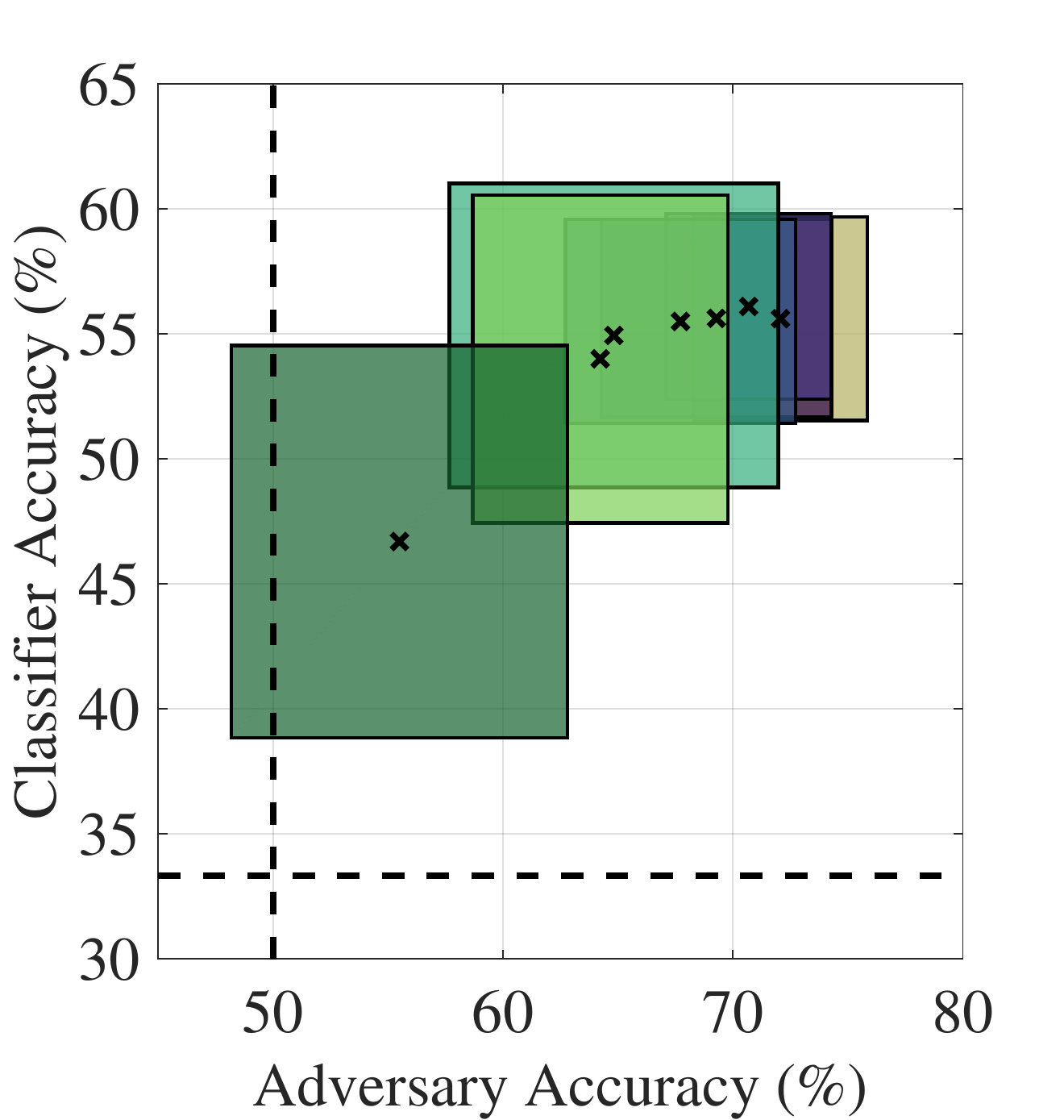}}
    \caption{Classifier versus adversary accuracies on the test sets of each subject after model learning. For each colored box, center marks denote the means across 5 training folds and 10 repetitions, and box widths denote $\pm$1 standard deviation intervals in both dimensions. Black dashed lines denote the change level accuracies in both axes. Note that the classifier handles a 3-class decoding problem, whereas the adversary considers a binary classification problem.}
    \label{fig:ss_plots}
\end{figure*}

Our main goal is to classify different textured surfaces that vary in their roughness levels based on EEG recordings during active tactile exploration. While we choose to exploit a convolutional neural network model for feature learning and classification, we further benefit from the promise of such models in learning nuisance-invariant discriminative representations. Figure~\ref{fig:ss_plots} presents an overview of subject-specific classification accuracies of the classifier and adversary networks on the test sets after model learning. Horizontal axes in Figure~\ref{fig:ss_plots} denote the binary adversary network classification accuracies for rub versus tap motor movement type classification, whereas the vertical axes denote the 3-class decoding accuracies for surface texture roughness. Center marks of each colored box denote the averages across 5 training folds and 10 repetitions. The width of the boxes in both dimensions denote $\pm{1}$ standard deviation intervals of the accuracies.

In cases where $\lambda=0$, the classifier is trained by the traditional categorical cross-entropy minimization. However we also learn an adversary network solely in parallel, which simply monitors the motor movement variant information leakage on the encoded latent representations $z$. The non-adversarial model ($\lambda=0$) 3-class classification results lead to an average $56\%$ accuracy across all subjects (maximum is $70.4\%$ for subject 7) as observed by the y-axis values in Figure~\ref{fig:ss_plots} plots. Simultaneously, we observed that indeed there is a significant amount of unwanted motor movement related information exploited in these texture roughness classification models. This is reflected by an average adversary network accuracy of $66.4\%$ (as high as $74.7\%$ for subject 2, x-axes in the Figure~\ref{fig:ss_plots} plots) when $\lambda=0$ for binary decoding of rub versus tap movement using simple linear classification with latent $z$. This supports our hypothesis on the necessity to censor motor movement relevant nuisance features from the latent representations during discriminative model learning for texture roughness classification. 

Our adversarial regularization approach as $\lambda>0$ successfully manipulates latent representations by reducing the adversary accuracies (i.e., leakage) towards minimizing the variance of distinct motor exploration movement patterns, while keeping the classifier performance stable for texture roughness decoding. On average across eight subjects, repetitions and folds, the $66.4\%$ adversary network leakage of regular model learning ($\lambda=0$) can be gradually reduced to $64.6\%$, $63.8\%$, $60.8\%$, $57.8\%$, $55.9\%$ and $52.9\%$ with adversarial regularization $\lambda$ values of $0.01, 0.02, 0.05, 0.1, 0.2, 0.5$ respectively. Such controllable invariance enables suppression of nuisance-related information down to near chance-level ($50\%$) decoding, while keeping classifier accuracies as stable as possible.

\begin{figure}
    \centering
    \includegraphics[width=0.6\textwidth]{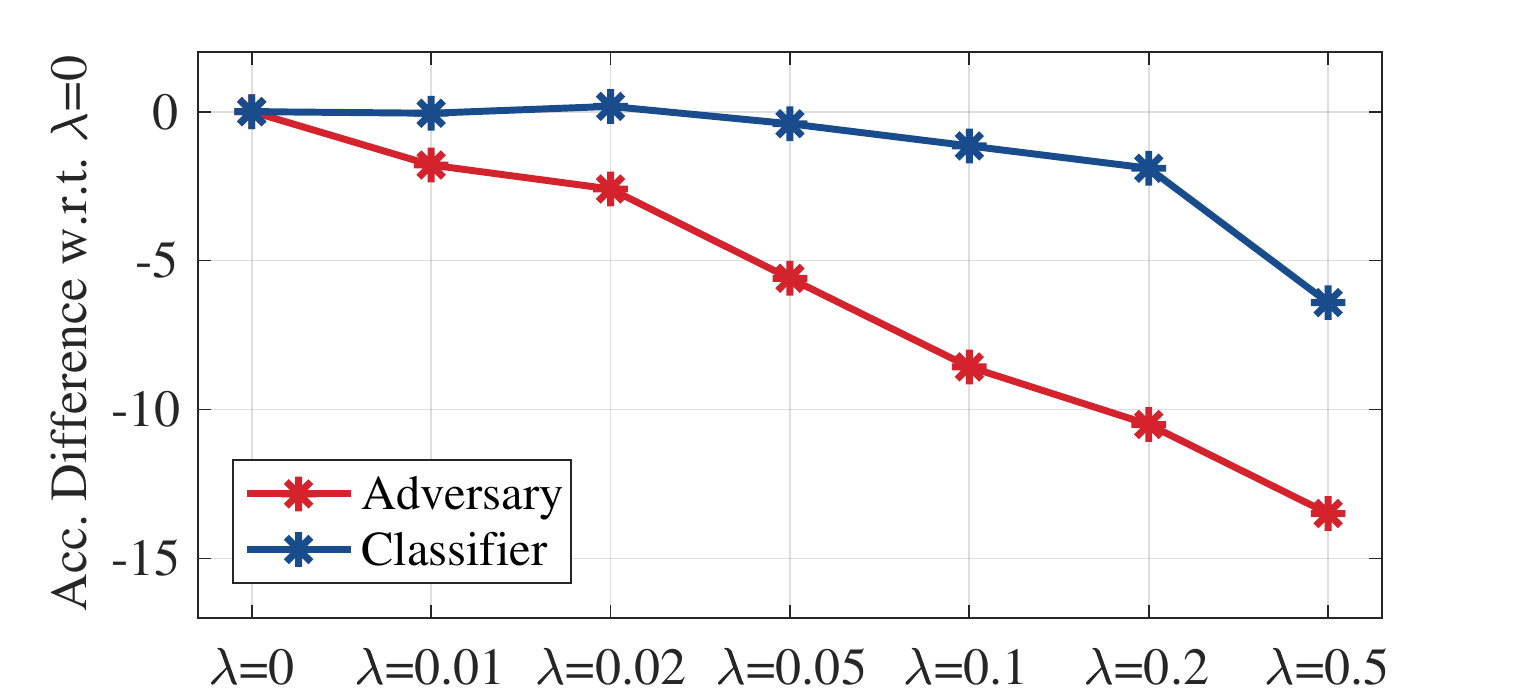}
    \caption{Difference of accuracies with respect to $\lambda=0$ (i.e., non-adversarial, regular CNN) for each adversarial regularization weight $\lambda$, averaged across subjects. Red denotes the differences in binary adversary network accuracies, and blue denotes the differences in 3-class classifier network accuracies.}
    \label{fig:diff_plot}
\end{figure}

Figure~\ref{fig:diff_plot} presents a summary of the differences between accuracies obtained with $\lambda>0$ models and the $\lambda=0$ condition. Horizontal axis represents the models for different $\lambda$ values, while the vertical axis represents the differences of accuracies averaged across subjects. Note that a very strong adversarial regularization can lead to losing class-discriminative information for the main classification task, hence hindering the texture roughness decoding accuracies (e.g., with $\lambda=0.5$). Here we present our results in an exploratory fashion, by demonstrating performances with all $\lambda$ values. For further model selection and online use, an optimal $\lambda$ value based on the validation set classifier and adversary accuracy can be selected \citep{ozdenizci2020learning}.

\section{Discussion}

The ultimate goal of this work is to build an EEG-guided system that can provide the sensation of objects with different surface roughness levels. During our daily activities we engage in active exploration of our surroundings, and neural responses to this type of exploration reflect both motor movement and sensory perception related information. The human brain processes motor and sensory related information in a coupled way, where both information are in affect with each other \citep{engel2013s,konig2013predictions,gallese2005brain,melnik2017eeg}.
Moreover, this active tactile exploration provides substantial information about the surface properties and demonstrate better discrimination thresholds compared to passive exploration \citep{lederman2009haptic,gibson1962observations,hollins2000evidence}. To that extent, we highlight the need to investigate discriminative models that are sensitive to sensory information while at the same time invariant to varying motor movement related cortical activity patterns.

In order to empirically assess the feasibility of this problem, we present the current experimental active tactile exploration study with healthy subjects while the participants use their index finger to either rub or tap surfaces with three different texture roughness levels. We propose to use an adversarial invariant representation learning neural network model to classify the three different texture roughness levels using $\sim$330 ms EEG data segments recorded in response to these movements, while minimizing the discriminability of the motor movement type. Our results show that the proposed approach can discriminate between three different textured surfaces with accuracies up to 70\%, hypothetically by exploiting event-related potentials generated in response to sensory information \citep{ballesteros2009erp}. Furthermore we also demonstrate the inherent existence of undesired motor movement related activity embedded in the predictive EEG features (as presented in Figure~\ref{fig:ss_plots}), pointing to the necessity of our approach in censoring motor movement relevant variability from the texture roughness discriminative representations during model learning. Overall, our proposed adversarial inference approach successfully demonstrates its capability to manipulate latent representations towards minimizing the discriminability of distinct motor exploration movement patterns, while keeping the classifier performance stable for texture roughness decoding.

Natural characteristics of EEG signals are particularly an important factor that motivates: (1) our adversarially learned neural network invariance approach to this complex neural decoding problem, and (2) our specific spatio-temporal feature learning network architecture choices. Our basis argument regarding (1) arises from the fact that EEG signals, by nature, are collected and analyzed as a superposition of multiple task-specific and task-irrelevant neural oscillations \citep{niedermeyer2005}. Since multi-channel time-series EEG signals do not trivially dissociate multiple cortical processes occurring at the same time, this characteristic feature took a significant role in motivating the invariant representation learning concept of our model. Our adversarial invariance learning setting is one approach against this phenomena towards disentangling specifically defined nuisance variability, which in this case is simultaneously present motor cortical de-/modulations. In different settings this invariance could address another a priori defined nuisance variability (e.g., rhythmic eye-blinks). Our choices regarding (2) simply indicate that the specific recording and pre-processing pipeline of EEG signals constitute deterministic choices for the network architecture. Our decision to use a generic convolutional network architecture based on EEGNet \citep{lawhern2018eegnet} is motivated by an approach to initially perform temporal convolutions analogous to frequency filtering activities for multi-channel EEG signals, spatial convolutions for localization of task-discriminative cortical sources/regions for each subject, followed by spatio-temporal convolution layers for summarization of extracted temporal and spatial features. Furthermore, as outlined in Section~\ref{sec:implementation}, our temporal and spatial convolution kernel sizes were chosen in consistency with the utilized EEG signal's recording and pre-processing pipeline (e.g., temporal convolution kernels were determined as one fourth of the EEG sampling rate to sufficiently capture periodic signal activity above 4 Hz). Hence the EEG signal characteristics implicitly determines the proposed model specifications in our study.

Our use of an invariant representation learning network offers the flexibility to manipulate the learned representations during training to be invariant to particular pre-defined nuisance variability (i.e., active touch movement condition), leading to significantly reduced motor movement related information leakage from the learned texture-discriminative representations. However one important limitation of our approach is that a very strong adversarial regularization can lead to loss of class-discriminative information for the main classification task, hence hindering the texture roughness decoding accuracies (as illustrated in Figure~\ref{fig:diff_plot} with $\lambda=0.5$). An immediate potential improvement to our current approach will be to address the systematic subject-specific model selection issue for the optimal $\lambda$ choice, which yields non-confounding results in terms of task discriminative information loss. Another important limitation of our study is the lack of subject-specific multi-session decoding stability analyses, due to our single-session experimental study design. To proceed along this line, a methodological improvement towards learning both motor movement as well as session-invariant representations can be potentially studied.


\section{Conclusion}

The overarching goal of this work is to build an EEG guided system that could provide and mimic the sensation of objects with varying levels of roughness. Applications of this system can include teleoperation, training of physicians training in virtual environments or remote robotic prosthesis control. Our daily life activities engage active exploration of our surroundings, and EEG responses to this type of exploration reflects both motor movement related and sensory perception information. Hence, we shed light onto the need to investigate discriminative models that are sensitive to sensory information while at the same time invariant to varying motor movement related cortical activity patterns.

In this paper, we show that it is possible to discriminate surfaces with varying levels of roughness during active tactile exploration using EEG as an input to the proposed neural network architecture. Results show that the proposed approach can discriminate between three different textured surfaces with accuracies up to 70\%, while suppressing movement related variability from learned representations. In summary, our aim of this study was to develop a methodology to extract and select EEG features that can classify various textures with different levels of roughness using EEG that are invariant to motor activity. Such EEG features will be used to design principles for model-based optimal EEG-guided haptic feedback system in our future work.

\section*{Acknowledgments}

Our work was supported by NSF grants IIS-1717654, IIS-1715858, IIS-1915083, IIS-1844885, CPS-1544895, CBET-1804550, M3X-20040457, and NIH grant 2R01DC009834.


\end{document}